# Strain release at the graphene-Ni(100) interface investigated by *in-situ* and *operando* scanning tunnelling microscopy

*Zhiyu Zou[1], Laerte L. Patera[1,2,†], Giovanni Comelli[1,2], Cristina Africh[1,*]*

[1]CNR-IOM, Laboratorio TASC, S.S. 14 km 163.5 - Basovizza, 34149 Trieste, Italy.

[2]Department of Physics, University of Trieste, via A. Valerio 2, 34127 Trieste, Italy.

---

[1] Corresponding Authors. Email: * africh@iom.cnr.it (Cristina Africh)

† Present Address: Department of Chemistry, Technical University of Munich, 85748 Garching, Germany.

**ABSTRACT**

Interface strain can significantly influence the mechanical, electronic and magnetic properties of low-dimensional materials. Here we investigated by scanning tunneling microscopy how the stress introduced by a mismatched interface affects the structure of a growing graphene (Gr) layer on a Ni(100) surface in real time during the process. Strain release appears to be the main factor governing morphology, with the interplay of two simultaneous driving forces: on the one side the need to obtain two-dimensional best registry with the substrate, via formation of moiré patterns, on the other side the requirement of optimal one-dimensional in-plane matching with the transforming nickel carbide layer, achieved by local rotation of the growing Gr flake. Our work suggests the possibility of tuning the local properties of two-dimensional films at the nanoscale through exploitation of strain at a one-dimensional interface.

**1. Introduction**

The capability of the graphene (Gr) lattice to sustain an enormous strain[1] allows for tuning its electronic and mechanical properties by applying external forces[2-3]. Previous studies show that strain in Gr lattice can induce band-gap opening[4-5], enhanced electron-phonon coupling [6] and strong pseudo-magnetic field [7], paving the way to the integration of various functionalities in graphene-based devices.

Chemical vapor deposition (CVD) is one of the most efficient methods for synthesizing wafer-sized single crystals of Gr and two-dimensional (2D) semiconductors [8-13]. At the graphene-metal interface, strain typically arises due to the lattice mismatch and the different thermal expansion coefficients between Gr and the substrate [3,5,14]. Lattice strain can influence the morphology and the band structure of Gr even at the nanoscale [5,15]. For example, in Gr nanobubbles formed on Pt(111) surfaces, Gr lattice undergoes extreme strain which induces a large pseudo-magnetic field [7]. Schumacher *et al.* reported that the strain in Gr grown on Ir(111) can influence the intercalation of Eu atoms underneath [16]. Boukhvalov *et al.* showed by simulation an enhanced chemical reactivity in highly corrugated Gr towards molecular hydrogen functionalization [17]. Recently, lattice strain has been found at one-dimensional mismatched interfaces of the lateral heterojunctions composed of two-dimensional layered semiconductors, showing that local lattice distortions influence the band structure of the heterojunction [18]. However, it should be noted that to date most of the reported studies were obtained through post-growth analysis, and little is known about how the growth mechanisms of 2D atomic crystals are influenced by such mismatched interfaces.

Here we report an *operando* investigation of Gr growth at the one-dimensional boundary with surface nickel carbide ($Ni_2C$) on Ni(100) substrates, both in the form of single crystals and oriented grains of a polycrystalline foil. Atomic-scale growth mechanisms have been unveiled by *in-situ* scanning tunneling microscopy (STM) performed at elevated temperatures (400-550 °C). By using the moiré superstructure as a magnifying glass, even small in-plane distortions due to the strain caused by the lattice mismatch at the $Ni_2C$-Gr interface can be visualized, and the dynamic influence of strain on the structure of growing Gr is revealed by means of real-time STM imaging performed at elevated temperatures. It is generally agreed that the structure of graphene grown on transition metal surfaces via the CVD method is determined at the nucleation stage [19-20] and is also driven by the strain release of the 2D coincidence lattice of the moiré superstructures [21]. Our work shows that, besides these aforementioned factors, the orientation of graphene is also affected during the expansion stage by the strain at the one-dimensional (1D) in-plane interface. This is an elusive effect that cannot be captured by post-growth measurements, but is relevant in the selection of catalysts and experimental conditions for the controlled growth of large single crystals of graphene and other 2D materials.

## 2. Methods

Polycrystalline Ni foils were pretreated following the procedures described in our previous publication to make the surface suitable for STM characterization [22]. STM measurements were conducted in a two-chamber system (base pressure: $1\times10^{-10}$ mbar) equipped with preparation and characterization (STM & LEED) facilities. Before Gr growth, the samples, both Ni(100) single crystal and polycrystalline Ni foils, were cleaned by several sputtering and annealing cycles in the ultrahigh vacuum (UHV) chamber. Gr was always grown when there were no contaminants detected by STM. Experiments on single crystals and on oriented grains of polycrystalline foils yielded comparable results, showing that in the present case conclusions derived from model system investigations apply also to more realistic substrates. Gr was grown at elevated temperatures (400-550 °C), either by backfilling the chamber with ethylene ($p$ = $5\times10^{-8}$-$5\times10^{-6}$ mbar) or via segregation from dissolved carbon, while the process was monitored in real time by continuous STM imaging. Both *operando* monitoring and post-growth measurement were performed by means of an Omicron variable temperature STM (VT-STM). Comparable results were obtained by using the two alternative growing methods. Besides conventional STM frame acquisition, an add-on FAST module was used, which can increase the frame rate up to 100 Hz [23-24]. Scanning with conventional frame acquisition was performed in constant current mode, while Fast-STM scans were acquired in quasi-constant height mode. Raw Fast-STM data were processed by a custom-made Python package; single frames were then processed by Gwyddion [25].

## 3. Results and discussion

### 3.1 Atomic structure of the Ni$_2$C-graphene in-plane interface

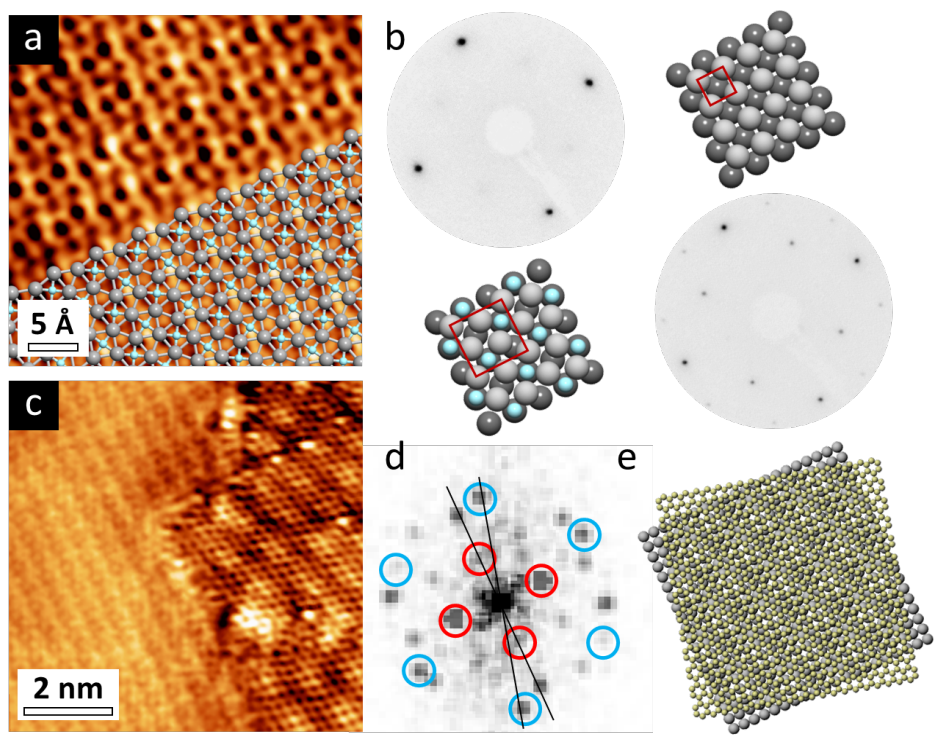

Figure 1. Structural identification of Ni$_2$C and in-plane Ni$_2$C-Gr boundary. (a) STM topographic image of surface carbide (Ni$_2$C) on Ni(100) ($V$ = -3 mV, $I$ = 30 nA), superposed by a ball-and-stick model of its atomic structure. (b) LEED patterns acquired at 70 eV and atomic models of clean Ni(100) (top) and Ni$_2$C (bottom) surfaces, respectively. The unit cells of the corresponding surfaces are denoted by red squares. (c) STM image taken at the Ni$_2$C-Gr boundary ($V$ = 100 mV, $I$ = 2 nA). (d) FFT of (c), exhibiting Ni$_2$C (in red circles) and Gr (in blue circles) spots. The black lines indicate the misorientation angle between Gr and the substrate [22]. (e) Visualization of the moiré superstructure observed in (c) by overlapping the Gr and surface Ni(100) lattices. Light grey/dark grey/cyan/yellow spheres represent Ni in the topmost layer/Ni in the second layer/carbon in carbide/carbon in Gr, respectively. (A colour version of this figure can be viewed online.)

Gr was synthesized, on both Ni(100) single crystals and polycrystalline Ni foils exhibiting micro-sized (100) grains, via CVD in a UHV chamber (see Methods and refs. [22,26] for details). Within a temperature range between 400 and 550°C, monolayer Gr growth is accompanied by the decomposition of a surface $Ni_2C$ phase [27-28]. Figure 1a shows the well-ordered $Ni_2C$ structure exhibiting a (2×2)$p4g$ symmetry [29-30], where nickel atoms are identified as depressions in STM (refer to the superimposed model). The surface reconstruction is also confirmed by the low energy electron diffraction (LEED) patterns (Figure 1b). While the LEED pattern of the clean Ni(100) surface exhibits only the spots of the (1×1) structure (top, atomic model in the right), the carbide one, with characteristic missing spots at normal incidence (bottom, atomic model in the left), indicates the (2×2)$p4g$ reconstruction [31].

Figure 1c shows an STM image acquired at the in-plane $Ni_2C$-Gr interface, which is typically aligned with the crystallographic directions of the Ni(100) substrate (see also Figure S4). On the left side, the characteristic (2×2) superstructure for $Ni_2C$ is clearly visible (note that here the Ni atoms appear as protrusions, due to a different tip termination). The right part of the image is covered by Gr, as evidenced by the rhombic moiré pattern [22]. Figure 1d shows the Fast Fourier Transform (FFT) patterns of the $Ni_2C$ (red) and Gr (blue) from Figure 1c. The relative rotation angle between the lattices of Gr and underlying Ni(100) surface, i.e. the misorientation angle $\theta$ [22], can be determined as 14.8°, given that the reciprocal vectors of the surface carbide supercells and of the Ni (100) surface are aligned. Overlapping the lattices of Gr and Ni(100) surface with the deduced misorientation angle, the moiré superstructure in Figure 1c can be easily reproduced (Figure 1e).

In Figure 1c, the seamless stitching at the border by C-Ni chemical bonding indicates that Gr lies in the same atomic layer of $Ni_2C$, as previously observed for the Gr growth on Ni(111) [28] and at variance with the cases of early transition metals (such as Mo, W), where Gr grows as an adlayer on metal carbides [12,32]. Under various STM imaging conditions, the $Ni_2C$ layer has an average apparent height larger than that of Gr (see Figures S1,2). According to our previous work, the adsorption height of Gr moirés on Ni(100) surface (the distance between the topmost Ni layer underneath Gr and the carbon atoms in the lowest valley of Gr moirés) is 1.95 Å, and the corrugation of moiré stripes can be up to 1 Å (see also Figure S3) [22]. For surface $Ni_2C$ on Ni(100), the carbon atoms sits 0.1 Å above the surface Ni atoms, and 1.97 Å above the second layer Ni atoms [33]. Therefore, on average the geometric height of Gr is larger than that of $Ni_2C$. This indicates that the apparent STM height difference between Gr and $Ni_2C$ is not a pure geometrical effect, but depends significantly on the electronic structure of both surfaces.

**3.2 Quenching the rippling of graphene moiré by the interface strain**

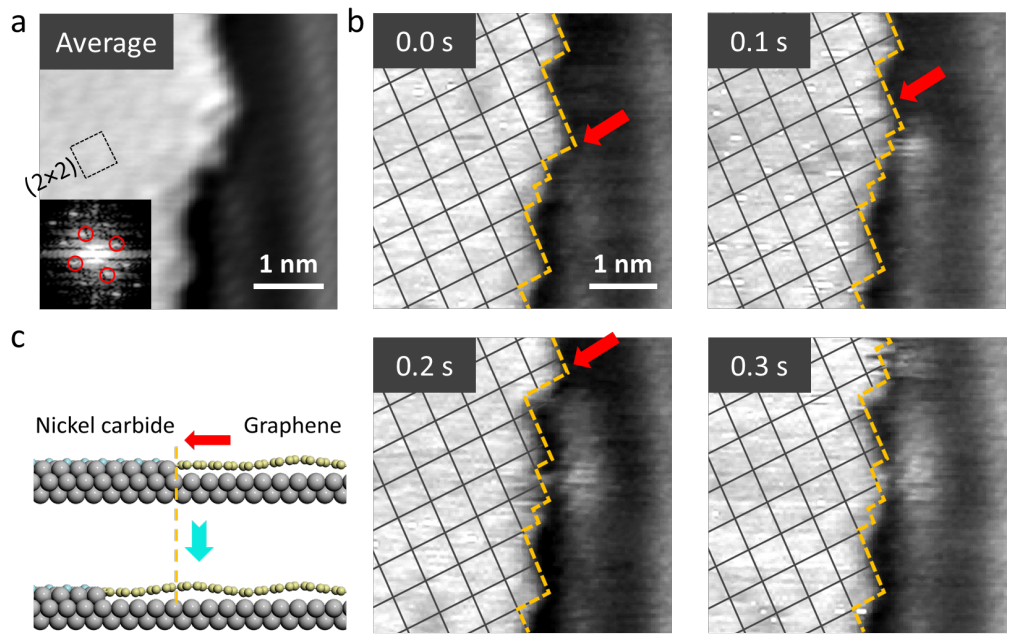

Figure 2. In-plane conversion at the $Ni_2C$-Gr interface imaged by high-speed STM (frame rate = 10 Hz, $V$ = -10 mV, $I$ = 13 nA, $T$ = 450 °C). (a) An average of 60 consecutive frames exhibiting the (2×2) superlattice of $Ni_2C$ (highlighted by the square) and the moiré stripes of Gr (on the right side, vertical). The depressions corresponding to Ni atoms are elongated in a specific direction probably due to the clock- or anti-clockwise rotation of Ni atoms. Inset: Fourier transform of the nickel carbide part of the image with the red circles highlighting the (2×2) superstructure. (b) STM sequence acquired in quasi-constant height mode. The grids denote the (2×2) superlattice of nickel carbide. Thermal drift can be neglected due to its careful compensation during the real-time measurements as well as the short timescale corresponding to each frame (0.1 s). The slight distortion of the square lattice can be attributed to the asymmetric lateral response of the STM scanner to the applied voltage. The yellow dashed lines denote the $Ni_2C$-Gr boundaries. The misorientation angle $\theta$ is ~3.5° (see details for the calculation of the misorientation angle in Section 3 of the SI). The red arrows point to nickel atoms that are going to be removed. (c) Schematic illustration of the in-plane Gr growth process at the interface. Grey/cyan/yellow spheres represent Ni/carbon in carbide/carbon in Gr, respectively. (A colour version of this figure can be viewed online.)

To clarify the Gr growth mechanisms at the mismatched interface, the in-plane conversion from carbide to Gr was imaged in real time by STM at both standard (Figure S4) and fast scan rates (~0.02 and 10 Hz, respectively) [23,34]. Both the (100) facet in polycrystalline Ni foils and the single crystal surfaces exhibit similar growth mechanism (see also the discussion in Section 2 of the Supporting Information). In the fast-scanning measurements (Figures 2a,b and Movie S1), nickel carbide (left) appears brighter (i.e. higher) than Gr (right), consistent with the STM results obtained in constant-current mode with a normal scan speed at both elevated and room temperatures (Figures S1,2,4). The average of 60 consecutive frames, shown in Figure 2a, unambiguously exhibits the (2×2) superlattice of $Ni_2C$ (see also the Fourier transform in the inset) [29]. Exploiting high-speed imaging, the removal of a few or even individual Ni atoms from the carbide layer (marked by red arrows), accompanied by the growth of Gr, can be observed (Figure 2b). In this sequence, the moiré stripes are parallel to the overall direction of Ni$_2$C-Gr interface. Notably, in the first frame (0.0 s) the lower part of the left moiré stripe is brighter than the upper part, in contrast with the STM appearance of extended Gr domains grown on flat Ni(100) terraces, where the moiré stripes of ridges or valleys can extend for tens or hundreds of nanometers without a significant modulation of their apparent height [22]. Along the sequence, as the Ni atoms diffuse away from the interface, the bright portion of the moiré stripe extends (0.1 s and 0.2 s), until it takes the full image length (0.3 s). This indicates that at the beginning of the STM sequence part of the Gr moiré remains in a strained, flattened morphology, and the strain is released only when the ridge of the moiré rippling is far enough from the in-plane interface, as illustrated in Figure 2c. This can be understood by considering that the height of Ni atoms in the surface carbide with respect to the Ni layer underneath (1.87 Å, corresponding to the interplanar distance between the topmost and second Ni layer) and that of C atoms in the valleys of Gr moiré (1.95 Å when $\theta = 0°$) are comparable. C-Ni chemical bonding at the interface thus tends to maintain the Gr film at about the same height of interface Ni atoms, locally hindering the mechanism usually adopted for the release of the strain induced by Gr bonding with the substrate, via formation of moiré wiggles. As the growth proceeds and Ni atoms detach from the boundary, opening enough room for the moiré to extend, the accumulated strain is released.

**3.3 Distortion of graphene lattices manipulated by the interface strain**

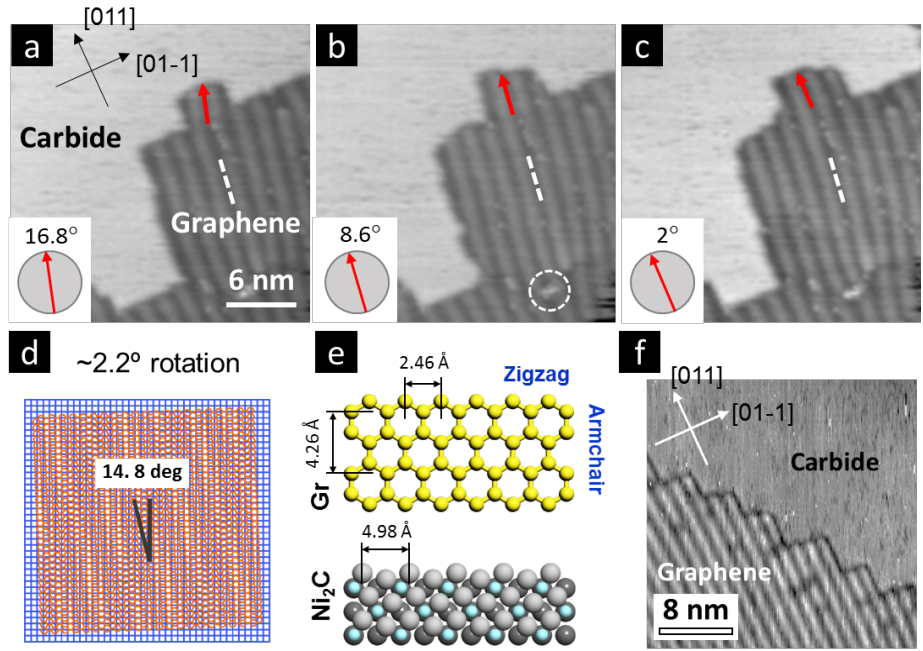

Figure 3. Shear force-induced rotation of Gr moiré stripes. (a-c) Three consecutive constant-current STM frames taken with a time interval of ~42 s at 465 °C ($V$ = -1.0 V, $I$ = 90 pA). The red arrows denote the orientations of the terminal moiré stripes; the angles are measured with respect to the close-packed direction of Ni(100). The defect highlighted by the white dashed circle was used for drift correction. (d) Amplification effect of the moiré superstructures on a small interplanar misorientation angle. (e) The configurations of a Gr ribbon surrounded by zigzag (top & bottom) and armchair (left & right) edges. The periodicities along the armchair and zigzag edges are indicated. Light grey/dark grey/cyan/yellow spheres: Ni in the topmost layer/Ni in the second layer/carbon in carbide/carbon in Gr. (f) Sawtooth-like in-plane interface when Gr is aligned with Ni(100). (A colour version of this figure can be viewed online.)

The interface strain can also be reflected locally by the orientation of Gr. As can be seen in Figure 3, in the interfacial regions where the moiré stripes are approximately orthogonal to the Ni$_2$C-Gr border, their orientation can change along the length of tens of nanometers, with a rotation angle that significantly varies during growth (see the red arrows in panels a to c). In the insets, we indicate for each panel the angle between the red arrow and the [011] Ni crystallographic direction (black arrow). It appears that, as the growth proceeds, the moiré directions tend to line up with the substrate: from a to c the orientation changes by 14.8°, until they become almost parallel. Considering that only the moiré stripes adjacent to the in-plane interface (red arrows) change their orientation, with other parts remaining undisturbed (white dashed lines), the observed effect cannot be due to thermal drift, which would result in a global misorientation. Therefore, we attribute this effect to a local change in the Gr lattice orientation. A simple sketch (illustrated in Figure 3d) shows that a 14.8° rotation of moiré stripes results from a ~2.2° misorientation angle of the Gr lattice with respect to the Ni close-packed direction. Thus, the linear moiré stripes can serve as a magnifying glass to reveal small changes of the Gr lattice, as previously reported [18].

Based on the geometric model of unstrained Gr on Ni(100) surface shown in Figure 3d, when the misorientation angle $\theta$ is small, the angular amplification factor $A_\theta$ can be expressed as:

$$A_\theta = \sin\varphi/\sin\theta \approx 1/(1-r)$$

Here $\varphi$ is the angle of the moiré stripes with respect to the crystallographic direction of the substrate and $r = l_{Gr}/l_{Ni}$, i.e. the ratio between lattice periodicities of Gr and Ni (100) (2.13 and 2.49 Å, respectively), yielding $A_\theta \sim 7$ (see Figure S5 for a detailed discussion).

A tentative explanation of the bending of Gr lattice during growth can be given taking into account the effect of the orientation of Gr on the interface atomic configurations. The lattice constant of nickel carbide (4.98 Å) is about twice the one of Gr (2.46 Å, see Figure 3e), indicating that for Gr aligned with the substrate (i.e. $\theta = 0°$), the zigzag edges can become commensurate with nickel carbide upon a small uniaxial strain (≈2%), thus maximizing at the same time both the in-plane C-Ni bonding and the interplanar chemical bonding with the Ni(100) substrate [22]. Therefore, when Gr is only slightly misoriented from 0°, there is always a driving force that tends to rotate it in order to align it with the substrate achieving lattice matching at the 1D interface, as observed in Figure 3a-c.

Such tendency is confirmed by Figure 3f, which shows a sawtooth-like $Ni_2C$-Gr interface. The moiré stripes are perfectly parallel to the [011] direction, and the interface along that direction, where Gr has zigzag edges, appears sharp and stable. In contrast, along the [01-1] direction, where Gr has armchair edges, the interface is not as straight. This again manifests that the interface of Gr with $\theta = 0°$ is favored, with stabilization mainly due to the zigzag edges, rather than the armchair edges, where the in-plane lattice mismatch with $Ni_2C$ cannot be easily accommodated (Figure 3e). For the armchair edges, presence of the Gr moiré wiggles further complicates the bonding with the flat $Ni_2C$ edge, introducing additional strain at the interface.

## 3.4 Discussion

The embedded growth of Gr in a metallic catalyst has previously been reported for both the Ni(111) and Cu(111) surfaces [28,35]. On Ni(100), Gr shows a similar growth mechanism, characterized though by more complicated interface structures. On the one hand, at variance with the (111) surfaces, the symmetry mismatch between Gr and Ni(100) excludes the possibility of an epitaxial interplanar registry, implying that the orientation of Gr has a distribution ranging from 0° to 15° [22]. On the other hand, the in-plane interface between Gr and Ni(100) imposes further constraints, (as shown in Figures 1,3), which also contribute to the determination of the Gr structure. The final configuration of the Gr/Ni(100) system appears therefore to result from the interplay between two effects that work in parallel: while the interaction between the Gr sheet and the underlying Ni(100) layer imposes the formation of moiré stripes at specific angles, the requirement of minimizing the strain, that is established at the in-plane $Ni_2C$-Gr 1D interface during growth, forces the system to adopt specific directions.

According to previous research, the lattice orientation of a graphene domain relies on its interlayer or in-plane interaction with the metal catalysts, and is determined during the nucleation stage [19-20]. Since line-defects which can impair the electric performance of graphene films will form when two misoriented domains meet during the catalytic growth [36], it is of utter importance to control the orientation of graphene nuclei in the CVD preparation to obtain a single crystalline film [37]. However, our work shows that the orientation of graphene domains with a large size can still be steered by the in-plane interface at the expansion stage, due to the lattice mismatch at the interface. Defects can thus be produced when joining graphene patches characterized by different distortions introduced by the mismatched in-plane interface, indicating that controlling the growth nucleation stage could be insufficient to obtain a graphene single crystal.

It has been reported that spatially periodic lattice potential introduced by the moiré superstructures can lead to novel band structures of Gr, such as the superlattice Dirac points and minigaps [38-40]. Therefore, the modification of periodicity or misorientation angle of the moiré superlattice by the interface stresses (as shown by Figures 3a-c) will mediate the electronic properties of Gr depending on the local geometry between Gr and nickel carbide. Besides, the uniaxial tensile strain can break the Gr sublattice symmetry, and thus open a bandgap which is proportional to the strain [41]. Considering the strong chemical bonding between Ni and C atoms [42] at the in-plane interface, a substantial uniaxial strain can be exerted on Gr, as shown in Figure 2b, which could introduce a local bandgap opening in Gr. Such phenomena can shed light on the strain engineering of Gr at the nanoscale.

## 4. Conclusion

The in-plane growth mechanism of Gr on Ni(100) surface has been investigated via *in-situ* scanning probe microscopy, shedding light on the effects of lattice strain in determining the Gr rippling and orientation. During the growth, Gr is interfaced to nickel carbide island seamlessly with the atomic bonding, giving rise to substantial strain in the Gr lattice. According to the interface atomic structures, strain can induce flattening of Gr rippling or twisting of the lattice, highlighting the role of interface configurations at the nanoscale in determining the overall morphology of a growing 2D material.

In conclusion, exploiting the magnifying effect of the moiré superstructures, we have revealed in real time during growth the formation of small, temporary deformations induced by interface strain in 2D lattices, which drive the system to its final relaxed morphology. The observed behaviour may stimulate new studies to investigate the possibility of locally tuning the band structure and chemical reactivity of 2D materials at the nanoscale by quenching the growth process at intermediate stages, before the strain release at the 1D interface of the growing flake has taken place.

**Appendix: Supporting Information**

Analysis of STM images and detailed geometric model for moiré patterns (pdf file)

STM movie Movie_S1 (mp4 file)

**Author Contributions**

The manuscript was written through contributions of all authors. All authors have given approval to the final version of the manuscript. Zhiyu Zou: Conceptualization, Investigation, Visualization, Writing - original draft. Laerte L. Patera: Investigation, Writing - original draft. Giovanni Comelli: Supervision, Writing - review & editing. Cristina Africh: Supervision, Writing - review & editing, Funding acquisition.

**ACKNOWLEDGMENTS**

Z.Z. acknowledges support by the "ICTP TRIL Programme, Trieste, Italy" in the framework of the agreements with the Elettra and CNR-IOM laboratories. We acknowledge support from EU-H2020 Research and Innovation programme under grant agreement no. 654360 Nanoscience Foundries and Fine Analysis – Europe. We thank M. Peressi for fruitful discussions.